\documentclass{aa}
\makeatletter
\let\@ln@colored\@empty
\let\@ln@normal\@empty
\renewcommand{\linenumberfont}{}
\let\linenumbers\relax
\let\nolinenumbers\relax
\@ifpackageloaded{linenoaa}{%
    \PackageWarningNoLine{aa}{linenoaa package loaded but disabled}%
    \let\linenumbers\relax
    \let\nolinenumbers\relax
}{}
\makeatother  
\usepackage{natbib} 
\usepackage{color}
\usepackage{footmisc}  
\bibpunct{(}{)}{;}{a}{}{,}
\usepackage[backref]{hyperref}
\PassOptionsToPackage{hyphens}{url} 
\usepackage{graphicx}             
\usepackage{float}                
\usepackage{booktabs}             
\usepackage{lipsum}               
\usepackage{subcaption}           
\DeclareUnicodeCharacter{FF0C}{,} 
\usepackage{lscape}             
\usepackage{placeins}

\begin{document}

   \title{Imaging-spectroscopy diagnosis of the giant sloshing spiral in the Virgo cluster with the Einstein Probe Follow-up X-ray Telescope}
    \titlerunning{Virgo cluster with EP-FXT}
   
\author{X. Zheng\inst{1,2}
        \and S. Jia\inst{1}
        \and C. Li\inst{1}
        \and Y. Chen\inst{1}
        \and H. Yu\inst{2}
        \and H. Feng\inst{1}
        \and D. Xu\inst{3}
        \and A. Liu\inst{2,4,5}
        \and L. Song\inst{1}
        \and C. Liu\inst{1}
        \and F. Lu\inst{1}
        \and S. Zhang\inst{1}
        \and W. Yuan\inst{6}
        \and J. Sanders\inst{4}
        \and J. Wang\inst{7}
        \and T. Chen\inst{1}
        \and C. Cui\inst{6}
        \and W. Cui\inst{1}
        \and W. Feng\inst{1}
        \and N. Gao\inst{1}
        \and J. Guan\inst{1}
        \and D. Han\inst{1}
        \and D. Hou\inst{1}
        \and H. Hu\inst{6}
        \and M. Huang\inst{6}
        \and J. Huo\inst{1}
        \and C. Jin\inst{5,6,8}
        \and M. Li\inst{1}
        \and W. Li\inst{1}
        \and Y. Liu\inst{6}
        \and L. Luo\inst{1}
        \and J. Ma\inst{1}
        \and G. Ou\inst{9}
        \and H. Pan\inst{6}
        \and H. Wang\inst{1}
        \and J. Wang\inst{1}
        \and J. Wang\inst{1}
        \and Y. Wang\inst{1}
        \and J. Xu\inst{1}
        \and Y. Xu\inst{6}
        \and X. Yang\inst{1}
        \and Y. Yang\inst{1}
        \and H. Zhang\inst{9}
        \and J. Zhang\inst{1}
        \and M. Zhang\inst{6}
        \and Z. Zhang\inst{1}
        \and H. Zhao\inst{1}
        \and X. Zhao\inst{1}
        \and Z. Zhao\inst{1}
        \and P. Zhu\inst{1}
        \and Y. Zhu\inst{1}
}
\authorrunning{X. Zheng et al.}

\institute{
State Key Laboratory of Particle Astrophysics, Institute of High Energy Physics, Chinese Academy of Sciences, Beijing 100049, China  
           \and School of Physics and Astronomy, Beijing Normal University, Beijing 100875, China
           \and Department of Astronomy, Tsinghua University, Beijing 100084, China
           \and Max Planck Institute for Extraterrestrial Physics, Giessenbachstrasse 1, 85748 Garching, Germany
           \and Institute for Frontiers in Astronomy and Astrophysics, Beijing Normal University, Beijing 102206, China
           \and National Astronomical Observatories, Chinese Academy of Sciences, 20A Datun Road, Beijing 100101, People’s Republic of China
           \and Department of Astronomy, Xiamen University, Xiamen 361005, China
           \and School of Astronomy and Space Science, University of Chinese Academy of Sciences, 19A Yuquan Road, Beijing 100049, People’s Republic of China
           \and Computing Center, Institute of High Energy Physics, Chinese Academy of Sciences, Beijing 100049, China
           }

   \date{Received September 30, 20XX}

  \abstract 
   {
We performed deep X-ray observations of the Virgo cluster using the \textit{Einstein Probe} Follow-up X-ray Telescope (\textit{EP-FXT}) with a total exposure of 295 ks. 
Taking advantage of the large field of view (FoV) and low particle background of \textit{EP-FXT}, the image reveals a giant spiral feature connecting the cold fronts in the northwest and southeast, forming a coherent structure consistent with previous \textit{XMM-Newton} and \textit{Suzaku} findings.
Furthermore, we present two-dimensional maps of the temperature, metallicity, and entropy in the Virgo cluster that cover a FoV of approximately $28.5^{\prime}$. These maps clearly depict a spiral structure that is characterized by high density, low temperature, high metallicity, and low entropy.
These results support the scenario in which the spiral morphology originates from gas sloshing that is induced by a minor merger.
In addition, the temperatures measured with \textit{EP-FXT} are reasonably consistent with those obtained from \textit{XMM-Newton} within the uncertainties. }


   \maketitle

\section{Introduction}\label{sec:intro}
Galaxy clusters are composed of hundreds to thousands of galaxies that are bound by gravity. They are the largest building blocks of the cosmic web \citep{kravtsov2012formation, vikhlinin2014clusters}. The X-ray radiation from clusters predominantly arises from the intracluster medium (ICM), which is a hot, diffuse plasma with a wealth of complex structures, such as shocks, bubbles, and cold fronts that trace past and ongoing violent dynamic processes in the system \citep{markevitch2007shocks}. 
Cold fronts are distinguished from shock fronts. They are sharp interfaces that separate brighter, denser, and cooler gases from the other side with a continuous pressure across the interface \citep{owers2009high, de2010cold, ettori2013cold,ghizzardi2010cold}.

Cold fronts or even spectacular spiral structures have been widely identified in nearby clusters \citep{keshet2012spiral}. 
It is proposed that the cold front originates in a displacement of cold gas at the cluster core, triggered by gravitational perturbations, such as those induced by minor mergers \citep{ascasibar2006origin, zuhone2010stirring, rossetti2013beyond, markevitch2001nonhydrostatic, roediger2011gas, markevitch2007shocks}.
In relaxed galaxy clusters, minor mergers disturb the central gas, initiating sloshing motions that produce multiple cold fronts at various locations over time \citep{roediger2011gas, liu2018inside}. 
These sloshing motions can evolve into intricate spiral structures that are visible in X-ray images as concentric edges \citep{ascasibar2006origin, sonkamble2024cool}, for instance, A795 \citep{kadam2024sloshingA795}.
Alternative explanations for the sloshing-induced spiral include two-phase gas flows with differing velocities \citep{keshet2012spiral} and the rotational motion of an in-falling gas flow around a moving brightest cluster galaxy (BCG) due to the Coriolis force \citep{inoue2022origin}.

Despite the detection of a variety of cold fronts, those that manifest themselves as a clear and complete spiral structure are rare, mainly because current X-ray telescopes cannot easily reveal large-scale and low-contrast structures. 
So far, remarkable examples include those seen in A496, A795, A1775, A2029, A2052, A2566, A2657, Centaurus, NGC 7618/UGC 12491, Perseus, and Virgo \citep{walker2017there,fabian2006very,roediger2012gas,ghizzardi2014metal,kadam2024sloshingA795,botteon2024radio,machacek2023chandra,blanton2011very,paterno2013deep,sanders2016very,botteon2021nonthermal,kadam2024sloshingA2566}, 
which were mainly found in mosaic observations with
\textit{XMM-Newton}\footnote{The X-ray Multi-Mirror Mission} and \textit{Chandra}\footnote{Chandra X-ray Observatory}.
The detection and detailed investigations of large-scale sloshing spirals are essential for advancing our understanding of the physical mechanism of sloshing and of the dynamical evolution of galaxy clusters.

Because of its proximity, the Virgo cluster is an excellent target for in-depth studies on this topic. Within the Virgo cluster, M87 stands out as the largest and brightest elliptical galaxy, with a redshift of 0.0042. It has been extensively studied at multiple wavelengths, which provided valuable insights into the interaction between its central supermassive black hole ($3.2 \times 10^{9} M_{\odot}$ within 3.5 pc of its center) and the surrounding ICM \citep{harms1994hst,simionescu2010metal,bahcall1977parameters}. 
Early X-ray observations with \textit{ROSAT}\footnote{ROentgen SATellite}, \textit{XMM-Newton}, and \textit{Chandra} have provided definitive evidence that the feedback of an active galactic nucleus (AGN) surrounds the central galaxy M87 \citep{bohringer2001xmm,molendi2001metal,bohringer1995interaction,forman2003reflections,sanders2016detecting}.
In addition, several distinct structures within M87 have been discovered, such as two X-ray arms, bubbles, and cavities \citep{sanders2016detecting}.
X-ray observations have revealed multiple sloshing cold fronts in the Virgo cluster. \textit{XMM-Newton} identified cold fronts at 33 kpc and 90 kpc and revealed a spiral structure that extends to approximately 200 kpc \citep{simionescu2010metal,roediger2011gas}. \textit{Suzaku}\footnote{originally named Astro-E2} later detected additional outer cold fronts at 233 and 280 kpc and carried out detailed radial analyses of their thermodynamic properties, which further confirmed that gas sloshing extends well beyond the core of the cluster \citep{suzaku2017}. 
Although previous studies have mapped the two-dimensional thermodynamic properties of the ICM in the central region of M87 \cite{gatuzz2022measuring}, the temperature, entropy, and metallicity distributions in the outer spiral arm remain poorly characterized.
While \textit{eROSITA}\footnote{The extended Roentgen Survey and Imaging Telescope Array} is capable of covering the extended spiral structure in the outskirts of Virgo \citep{mccall2024srg}, its relatively short exposure time has limited a detailed analysis of the associated physical properties.
To more comprehensively investigate the role of gas sloshing and cold fronts in the outskirts of galaxy clusters, it is important to conduct observations with an X-ray telescope that offers a larger FoV and sufficient sensitivity.

Benefiting from its large FoV and low particle background, the Follow-up X-ray Telescope on board the \textit{Einstein Probe} (\textit{EP-FXT})\citep{yuan2022einstein,Yuan2025} is well suited for detecting large-scale extended structures with a low surface brightness.
We present deep \textit{EP-FXT} observations of the Virgo cluster.
Our data support the detection of the giant sloshing spiral and enable spatially resolved analyses of its thermodynamic properties.
The paper is organized as follows. We describe the observations in Sect.~\ref{sec:2}, and we present the imaging results in Sect.~\ref{sec:3} and the spatially resolved spectral results in Sect.~\ref{sec:4}. We discuss the physical implications in Sect.~\ref{sec:5} and summarize our findings in Sect.~\ref{sec:6}.

\section{Observations and data reduction}
\label{sec:2}

\textit{EP-FXT} is one of the two payloads on board the \textit{EP} mission, which is an X-ray telescope that is a result of China-Europe collaboration and was launched on January 9, 2024.
\textit{EP-FXT} is capable of conducting detailed observations of targets triggered by the \textit{Wide-field X-ray Telescope} (\textit{EP-WXT}) and of preselected targets designated from the ground.
\textit{EP-FXT} comprises two identical modules, FXT-A and FXT-B, that are each equipped with Wolter-I type X-ray optics featuring a focal length of 1.6 m and 54 nested gold-coated nickel mirror shells. These modules are paired with PNCCD detectors as focal-plane instruments and enable precise measurements of the photon energy, position, and timing \citep{cui2023design}. 
The device features a pixel array of 384 $\times$ 384, with an angular resolution of approximately $9.6^{\prime\prime}$ per pixel.
\textit{EP-FXT} covers an energy range of 0.3–10 keV, with a FoV of $1^\circ \times 1^\circ$, a spatial resolution of $22^{\prime\prime}$ in terms of half-power diameter (HPD), and an effective area of approximately 300 $cm^{2}$ at 1 keV (one unit). 
In addition, the particle background of \textit{EP-FXT} is about one-fifth of that of \textit{eROSITA}, and its low background characteristics make it particularly well suited for observing low-brightness diffuse emissions \citep{zhang2022estimate}.

As the first-light object and a calibration source for \textit{EP-FXT}, the Virgo cluster and its central galaxy M87 have been extensively observed. 
The large FoV of \textit{EP-FXT} means that a single observation can effectively cover the outskirts of M87.
All observations were performed in full-frame (FF) mode using either the thin or medium filter. 
Three of the FXT-B observations were conducted in closed mode, which rendered the corresponding data unusable. As a result, only FXT-A data were used. This resulted in a total effective exposure time of 295 ks
(see the observational log in Table~\ref{label1}).

\begin{figure}[h]
\centering
\includegraphics[width=\linewidth]{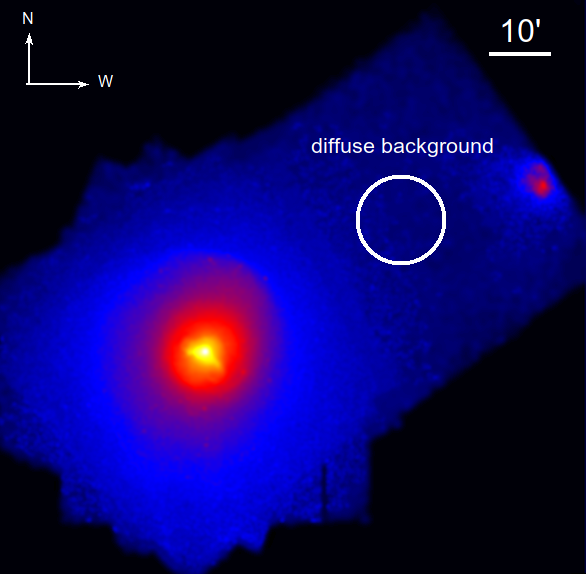}
\caption{Combined \textit{EP-FXT} X-ray image of the Virgo cluster in the 0.3-2.5 keV band. The image has been vignetting corrected, background and point-source subtracted, flattened, and adaptively smoothed. The white circle marks the background region. The bar has a length of 10$^\prime$. The arrows point north and west. 
}
\label{fig:full}
\end{figure}

We used the latest version 1.10 of the Follow-up X-ray Telescope Data Analysis Software (\texttt{FXTDAS})\footnote{\url{http://epfxt.ihep.ac.cn/analysis}} developed at the \textit{EP-FXT} Science Center for data reduction \citep{zhaohs2025submit}. New event files were generated through the \texttt{fxtchain} pipeline, and the out-of-time events were identified using \texttt{fxtootest}, which were then removed from the data with \texttt{ftimgcalc}\footnote{\url{https://heasarc.gsfc.nasa.gov/docs/software/lheasoft/help/heatools.html}}. When high-energy cosmic rays pass through space telescopes, they release energy onto the detectors, which excites fluorescence lines that are not modulated by the vignetting effect. To correct for this background contribution,
\texttt{fxtbkggen} was used to generate particle backgrounds for each observation.
We removed the particle background from each image and then used the \texttt{reproject\_image}\footnote{\label{ciao}Derived from the \textit{Chandra} Interactive Analysis of Observations (\texttt{CIAO}) version 4.16.0, available at \url{https://cxc.cfa.harvard.edu/ciao/}} tool to merge all exposure images along with their corresponding particle background-subtracted images.
Subsequently, we used the merged exposure maps to correct for the vignetting effect in the particle background-subtracted X-ray images.
Following this, the data from ObsID 13600005125, without particle background, were selected as the reference for the diffuse background.
Because the primary focus of this study is M87, we selected a background region in close proximity to M87.
This approach means that the diffuse background component surrounding M87 is probably not underestimated, thereby facilitating preciser background subtraction and strengthening the overall robustness of the analysis.
We combined multiple observations, subtracted the diffuse background, and normalized the image by the maximum exposure time to produce a calibrated X-ray image. 
We used \texttt{wavedetect}\footref{ciao} with scales = "4 8 16 32" and the default \texttt{sigthresh} to detect and remove point sources.
The voids were filled using \texttt{dmfilth}\footref{ciao} with the POISSON method and a random seed of zero to ensure reproducibility. 
Adaptive smoothing was then applied using \texttt{dmimgadapt} with Gaussian kernels, smoothing scales from 0.1 to 10 pixels, and a minimum of 30 counts per kernel.
The final X-ray image in the energy range of 0.3-2.5 keV is shown in Fig.~\ref{fig:full}.

Furthermore, to compare with previous results, we reanalyzed 
a \textit{Chandra} observation (ObsID 5826) with an exposure of 126 ks and 
an \textit{XMM-Newton} observation (ObsID 0803670601, EPIC-MOS2) with an exposure of 65 ks.
The data were reprocessed using the \textit{XMM-Newton} Science Analysis System (\texttt{SAS} \footnote{\url{https://www.cosmos.esa.int/web/xmm-newton/sas}}), 
and time intervals affected by flares were removed through high-energy light-curve screening.
In the subsequent \textit{XMM-Newton} analysis, blank-sky background files were employed to improve the accuracy of the background estimation and spectral analysis.

\begin{table*}[t]
\footnotesize
\caption{\textit{EP-FXT} observations of the Virgo cluster.}
\label{tab:obs}
\tabcolsep 11pt 
\begin{tabular*}{\textwidth}{cccccccc}
\hline\hline
   ObsID& RA & DEC & Date & Exposure time (ks) & filter & Instrument \\
  
    \hline
     08500000007 & 187.71 & 12.39 & 2024-02-22  & 3.38 & thin & A \\
     08500000008 & 187.71 & 12.39 & 2024-02-22  & 20.46 & medium & A \\
    13600005124 & 187.71 & 12.39 & 2024-03-17  & 77.73 & thin & A/B \\
    13600005125 (offset) & 187.01 & 12.87 & 2024-03-18 & 23.78 & thin & A/B \\
    13600005132 & 187.71 & 12.39 & 2024-04-05  & 28.08 & medium & A \\
    13600005431 & 187.71 & 12.39 &  2024-04-16 & 18.15 & medium & A/B \\
    13600006200 & 187.71 & 12.39 &  2024-05-20 & 97.17 & medium & A/B \\
    11904194486 & 187.71 & 12.39 &  2024-06-15 & 50.18 & thin \& medium& A/B \\
\bottomrule
\label{label1}
\end{tabular*}
\end{table*}

\section{Imaging analysis}\label{sec:3}

To better reveal structures in the image, we applied multiple image enhancement techniques to enhance the visibility of both sharp and diffuse edges, thereby highlighting structural details more effectively across different spatial scales.

\begin{figure}[h]
\centering
\includegraphics[width=\linewidth]{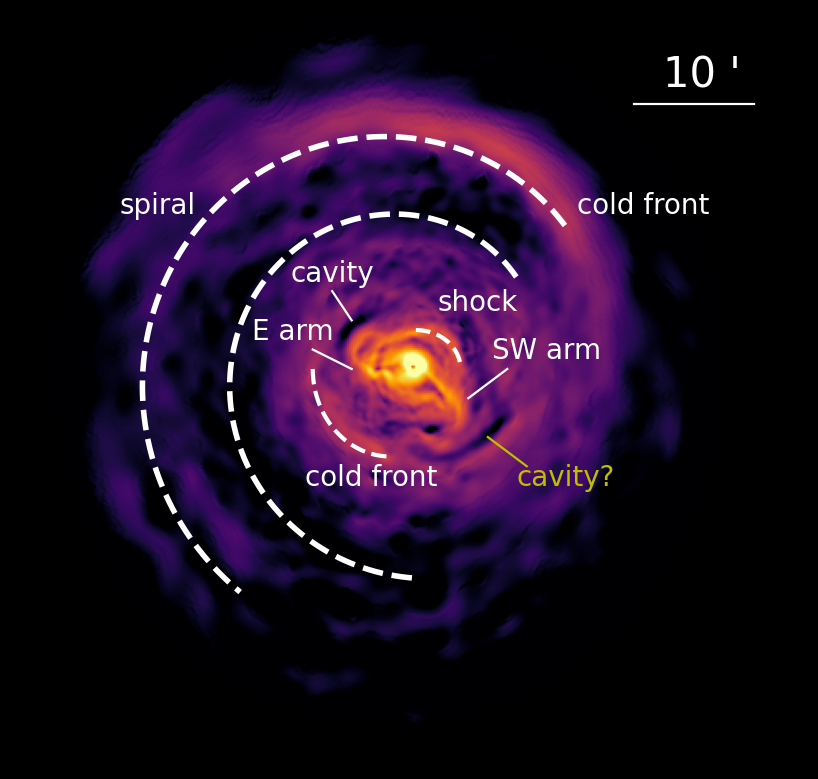}
\caption{Adaptive GGM filtered image in the central Virgo cluster in 0.3-2.5 keV with S/N = 70. Prominent features are labeled.  The two dashed lines that mark the giant spiral are identical to those in Fig.~\ref{fig:res}.}
\label{fig:ggm}
\end{figure}

Gradient filtering enhances the visibility of both sharp and flat image features, thereby facilitating their detection and analysis \citep{sanders2016very,sanders2016detecting,sanders2022studying}. 
Because of the high noise in X-ray image gradient measurements within low-brightness regions, the adaptive Gaussian gradient magnitude (GGM) filtering was employed to select the dynamic scale. This method dynamically adjusts filter parameters based on local gradients, enabling better capture of image textures and edges \citep{sanders2022studying}. Additionally, adjusting the signal-to-noise ratio (S/N) helps smooth the image, enhances the S/N, and reveals finer details. Using these techniques, we applied the adaptive GGM filter to the vignetting-corrected and background-subtracted image in the 0.3–2.5 keV energy range (see Fig.~\ref{fig:ggm}).

The image clearly shows two X-ray arms. The southwestern arm exhibits a clockwise rotation toward the southeast \citep{forman2007filaments}. This rotational feature may be attributed to the interaction between relativistic jets from the AGN and the ICM \citep{forman2003reflections}. Furthermore, the image also identifies a cavity, a shock to the northwest, and cold fronts to the southeast and northwest. These features are consistent with the analysis by \citet{sanders2016detecting} based on \textit{Chandra} data. We also identified a region in which the surface brightness was significantly reduced outward of the southwestern X-ray arm, which may indicate a cavity. 
The outskirts of M87 exhibit a prominent spiral morphology that extends from the northwest to the southeast, and similar spiral features are also visible in the \textit{XMM-Newton} mosaic image \citep{roediger2011gas,simionescu2010metal}.
We then generated a residual image to better visualize this large-scale faint feature.

\begin{figure}[h]
\centering
\includegraphics[width=\linewidth]{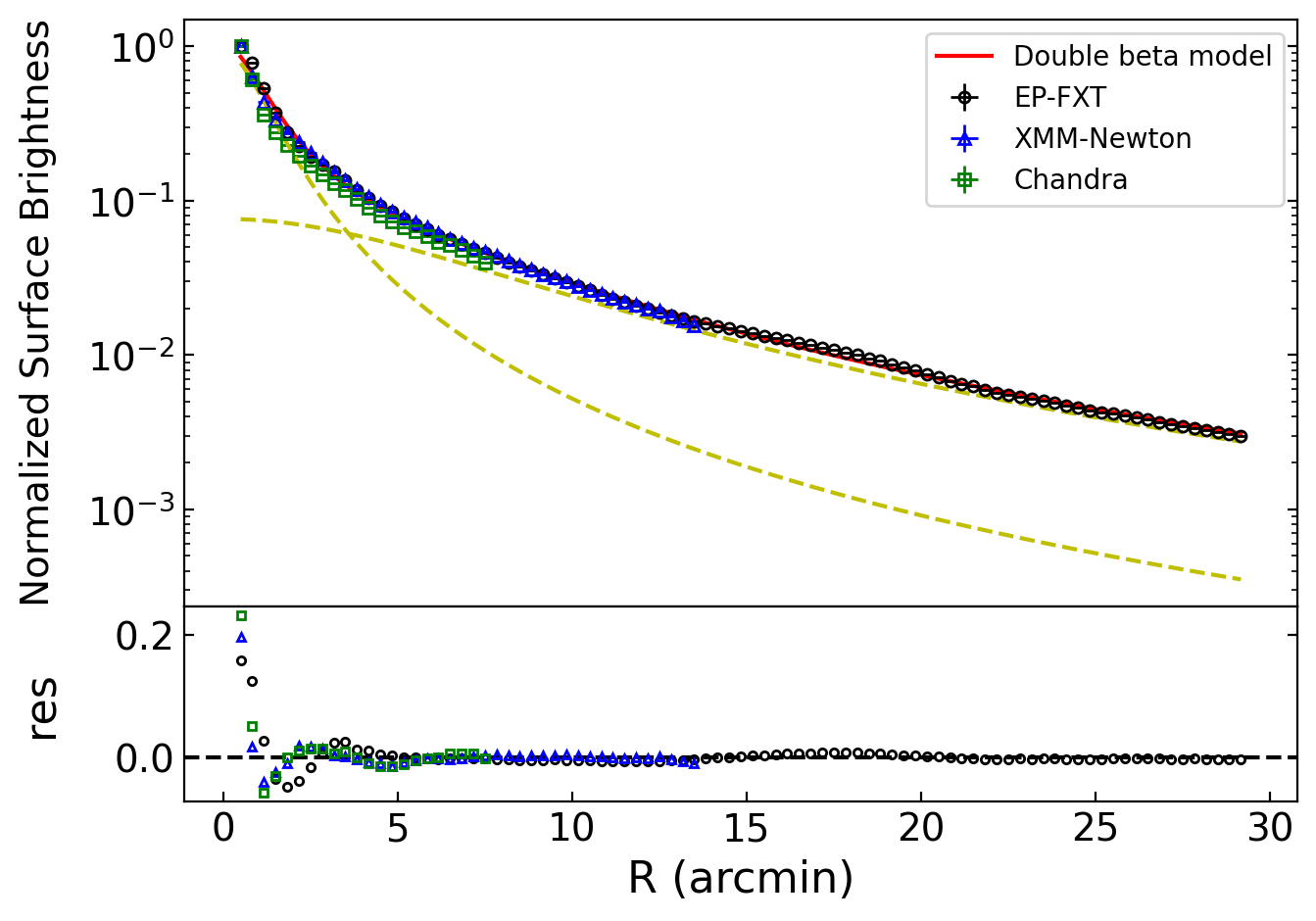}
\caption{Normalized surface brightness profiles measured with \textit{EP-FXT}, \textit{XMM-Newton}, and \textit{Chandra}. The \textit{EP-FXT} data were fit with a double $\beta$ model. Individual components are displayed as dashed yellow curves.}
\label{fig:1d_profile}
\end{figure}

First, we estimated the first-order radial profile using the well-known double $\beta$ model and subtracted it (see Fig.~\ref{fig:1d_profile}).  
Except for the central region that is affected by the AGN, the \textit{EP-FXT} data agree with the double $\beta$ model and are consistent with the radial profiles measured by the three telescopes. Furthermore, the \textit{EP-FXT}'s large FoV means that its observational range extends to regions up to $30^{\prime}$ from the outskirts of M87.

\begin{figure}[h]
\centering
\includegraphics[width=\linewidth]{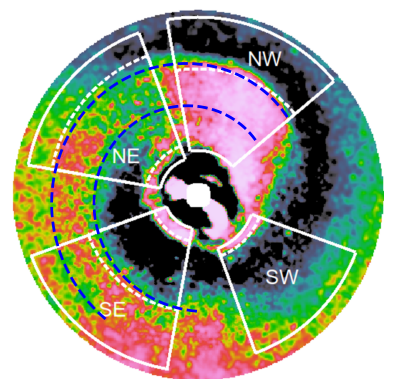}
\caption{Residual image obtained by subtracting the image in Fig.~\ref{fig:full} with the elliptical $\beta$ model. A Gaussian smoothing with a kernel size of 7 pix was applied (1 pixel = $9.6^{\prime\prime}$). Four regions of interest (NW, SE, SW and NE) are marked.  The dashed white lines mark the position of the radial discontinuity. The dashed blue lines mark the position of the giant spiral arm.}
\label{fig:res}
\end{figure}

To characterize the radial profile, we used Sherpa (version 4.16.0 within \texttt{CIAO})\footnote{\url{https://cxc.cfa.harvard.edu/sherpa/}} to model the surface brightness. 
To account for the apparently elongated morphology in the image, we fit an elliptical $\beta$ model to the data; the best‑fit parameters are listed in Table~\ref{tab:ell_sherpa}. We then subtracted this model from the original image to obtain the residual image shown in Fig.~\ref{fig:res}.

\begin{table*}[t]
\centering
\footnotesize
\caption{Best‐fit parameters of the elliptical $\beta$ model from Sherpa.}
\label{tab:ell_sherpa}
\begin{tabular}{ccccc}
\hline\hline
$r_{0}$ (arcsec) & $\mathrm{ellip}~(10^{-2})$ & $\theta$ (deg) & $A$ (counts pixel$^{-1}$) & $\alpha~(10^{-2})$ \\ 
\midrule
$90.58^{+0.39}_{-0.39}$ 
 & $10^{+0.04}_{-0.04}$ 
 & $248.13^{+0.12}_{-0.12}$ 
 & $4528.09^{+17.9}_{-17.9}$ 
 & $74^{+0.09}_{-0.09}$ \\
\bottomrule
\end{tabular}
\end{table*}

The residual image revealed a few patterns, including X-ray arms and discontinuities, which indicate cold fronts or shocks.
Notably, we identified a ring-like structure outside the X-ray arms that displays a break in the southern part.
This may be attributed to interactions between the AGN jets and the ICM \citep{forman2003reflections}.
Most importantly, a large-scale spiral structure with an excessive brightness is evident in the outskirts, where it spirals from the cold front in the northwest to the southern cold gas clump (marked with dashed blue lines in Fig.~\ref{fig:res} and dashed white lines in Fig.~\ref{fig:ggm}). 
Consistent with the mosaic \textit{XMM-Newton} results \citep{simionescu2010metal}, the \textit{EP-FXT} observations clearly reveal that the spiral features are continuous and well defined across the entire FoV, likely owing to the low particle background and wide spatial coverage of the instrument.

The outer northwestern edge (a cold front) of the spiral lies approximately $18.3^{\prime}$ away from the center. The structure rotates counterclockwise toward the south, and its inner southern edge is located approximately $17.5^{\prime}$ away from the center.

\begin{figure}[h]
\centering
\includegraphics[width=\linewidth]{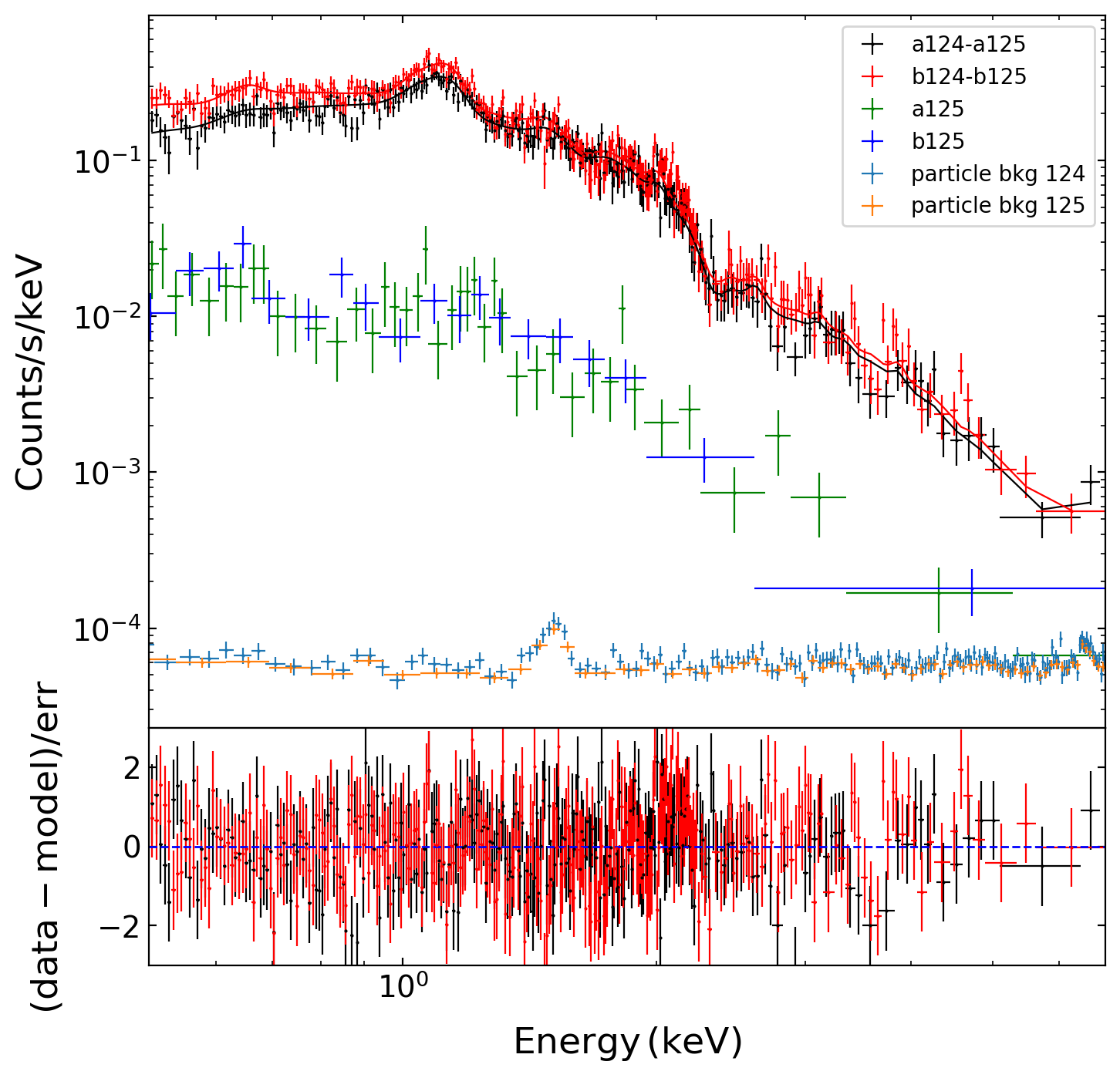}
\caption{
EP-FXT spectrum from a \texttt{contour-binning} region in M87. The black and red points represent the spectral data from FXT-A and FXT-B for ObsID 13600005124, with the corresponding background from ObsID 13600005125 subtracted, and the solid line represents the best-fit model.
The green and blue points represent the spectra from FXT-A and FXT-B for ObsID 13600005125, respectively. The light cyan and orange points represent the particle backgrounds for ObsIDs 13600005124 and 13600005125, respectively.}
\label{fig:spec}
\end{figure}

\begin{figure*}[th]
\centering
\includegraphics[width=0.49\linewidth]{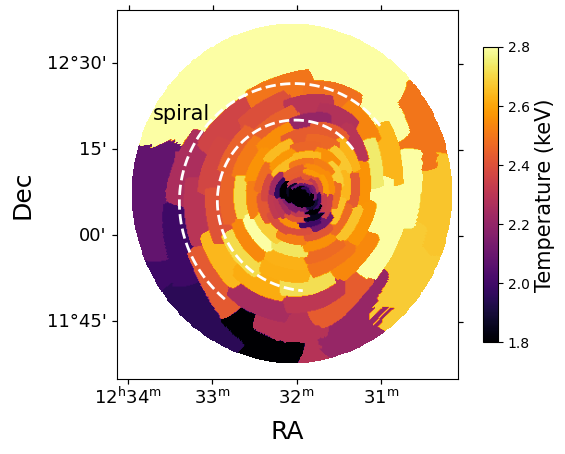}
\includegraphics[width=0.49\linewidth]{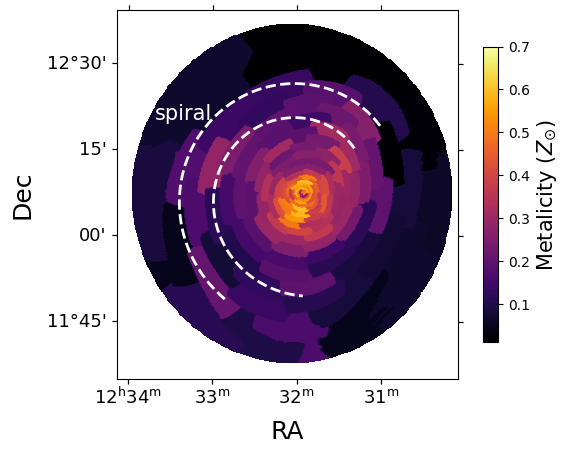}
\includegraphics[width=0.49\linewidth]{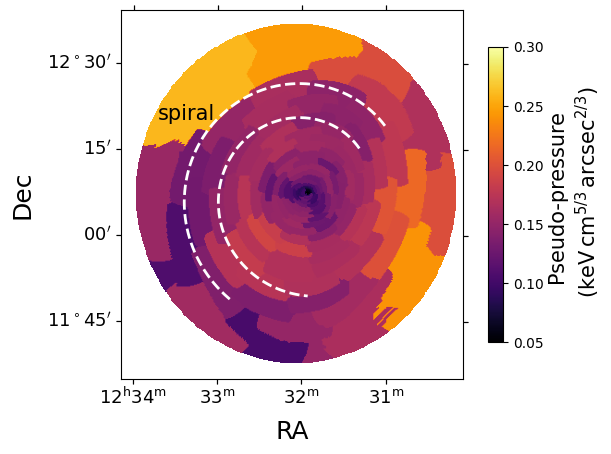}
\includegraphics[width=0.49\linewidth]{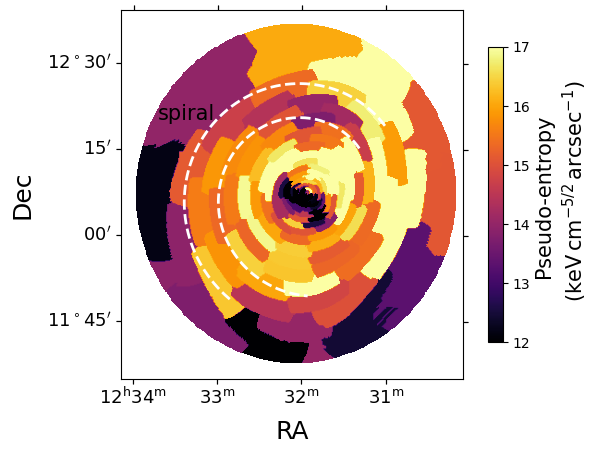}
\caption{Temperature, pressure, entropy, and metallicty maps measured with \textit{EP-FXT}, pixeled with the \texttt{contour-binning} method. The dashed white curves mark the spiral structure as in Fig.~\ref{fig:res}. 
}
\label{maps}
\end{figure*}

\section{Imaging spectral analysis}
\label{sec:4}
The X-ray image obtained with \textit{EP-FXT} was divided into 142 regions using the \texttt{contour-binning} software \citep{sanders2006contour} with a S/N of 300. 
We adopted a flat-weighting assumption, assuming that the X-ray emission within each region is spatially uniform, to calculate the ancillary response files (ARFs). The ARFs were generated using the \textit{EP-FXT} response tool \texttt{fxtarfgen}. During the calculation, the parameter \texttt{extend} was set to one to account for the extended nature of the source, thus more accurately capturing the spatial distribution of the emission within each extraction region.
The spectra were grouped using \texttt{grppha} with a minimum of ten counts per bin to ensure the applicability of the C-statistic.
To minimize the impact of background contamination on the source spectra and improve the accuracy of the spectral analysis, we selected observations of the M87 source and its corresponding background, ensuring similar observation times, the same filter settings, and longer exposure durations; specifically, ObsID 13600005124 and ObsID 13600005125.
In the spectral analysis, both the particle background and the diffuse background were taken into account.
We calculated the particle background for both observations, with a ratio of $1.05 \pm 0.03$, indicating that the particle background levels are nearly identical.
Therefore, we selected the spectra from the same CCD region in ObsID 13600005125 to serve as the background for ObsID 13600005124.
For regions in ObsID 13600005125 that were contaminated by bright point sources or extended sources, background data from nearby uncontaminated areas were used instead.
Because our analysis focuses on M87, we intentionally retained the Virgo cluster component in the background data because it reflects the diffuse emission environment surrounding the galaxy better.

We spectrally fit the combined FXT-A and FXT-B spectra in each region using XSPEC version 12.14.1 over the 0.5–7.0 keV energy range. 
The diffuse ICM thermal emission was described with the {\tt APEC} model (version 3.0.9), and the Galactic hydrogen absorption was accounted for using the {\tt TBABS} model, 
adopting the solar abundance table from \citet{anders1989abundances}, with the redshift fixed at 0.0042 and the hydrogen column density fixed at $2.11 \times 10^{20} \mathrm{cm}^{-2}$ \citep{willingale2013calibration}.
Other parameters were set to be free.
We examined the C-Stat/d.o.f. values for all regions, which were consistently around 1.2.
We plot the spectrum of one region in Fig.~\ref{fig:spec}.
The region belongs to the \texttt{contour-binning} partition and is located in a region approximately $6.6^{\prime}$ away from the cluster center.
The red and black points represent the spectral data from FXT‐A and FXT‐B in the source region (with the diffuse background subtracted), while the solid line denotes the best‐fit model. The C-Stat/d.o.f. for FXT-A and FXT-B are 246.74/287 and 295.70/299, respectively.
The green and blue points display the spectral data from FXT‐A and FXT‐B in the diffuse background region, which is co-located on the CCD with the source region. In addition, the light cyan and orange indicate the particle backgrounds for the source and diffuse background regions, respectively.

\subsection{The map of physical properties}

The temperature map (see Fig.~\ref{maps}) reveals a central region in M87 with temperatures below 2.5 keV, along with two distinct low-temperature regions in the northwest and southeast. Sharp temperature increases are seen outside these regions that are coincident with the locations of two concentric arc-shaped cold fronts.
A low-temperature belt extends outward from the central region. It connects the low-temperature regions in the southeast and northwest and spirals in counterclockwise from the northwest, forming a large-scale low-temperature spiral structure. 
In addition, a region with a lower temperature than the surrounding areas is located outside of the southwest arm.
A detailed analysis is presented in Sect. \ref{subsec 4.2}.

The metallicity was obtained from the {\tt APEC} model.
The metallicity is lower within the two X-ray arms, but higher north and south of the center, which is consistent with the findings by \citet{simionescu2007gaseous,gatuzz2022measuring}. Additionally, beyond $12^{\prime}$, the metallicity in the northwest is significantly higher than that in the southeast, likely due to the accumulation of cold gas in the northwest.
In addition, the metallicity of the spiral structure displays higher values, which further supports the existence of the spiral structure.

To investigate its thermodynamic properties, we examined the distributions of pseudo-pressure and pseudo-entropy. The thermal pressure and entropy can be defined \citep{sasaki2016x} as $ P = kTn_{\textnormal{e}}$ and $S = kTn_{\textnormal{e}}^{-2/3}$, respectively, where $n_{e}$ is the electron density, and $kT$ is the temperature. Based on the emission measure obtained from spectral fits, we obtained the pseudo-entropy and pseudo-pressure. We convert $P$ and $S$ into $P \propto kT{\rm norm}^{0.5}$ and $S \propto kT{\rm norm}^{-1/3}$, respectively \citep{fabian2006very}. 
Pseudo-pressure and pseudo-entropy maps are also displayed in Fig.~\ref{maps}.
In the central region of M87, the pressure map reveals decreasing pressure along the eastern and southwestern X-ray arms and increasing pressure in the northwest and southeast. The entropy map shows low-entropy features coincident with both X-ray arms, exhibiting an entropy asymmetry northwest and southeast near the core. These results are consistent with those reported by \citet{simionescu2007gaseous}. 
The outer spiral structure also displays lower entropy values, which further supports the existence of the spiral structure.

To summarize, the \textit{EP-FXT} observations revealed the spiral structure with high completeness and clarity. Moreover, we systematically mapped its thermodynamic properties, including temperature, metal abundance, and entropy, through two-dimensional distributions. These results provide detailed and spatially resolved evidence for the existence and evolution of the spiral structure.

\subsection{Radial sector profiles}
\label{subsec 4.2}

\begin{figure*}[tbh]
\centering
\includegraphics[width=0.49\linewidth]{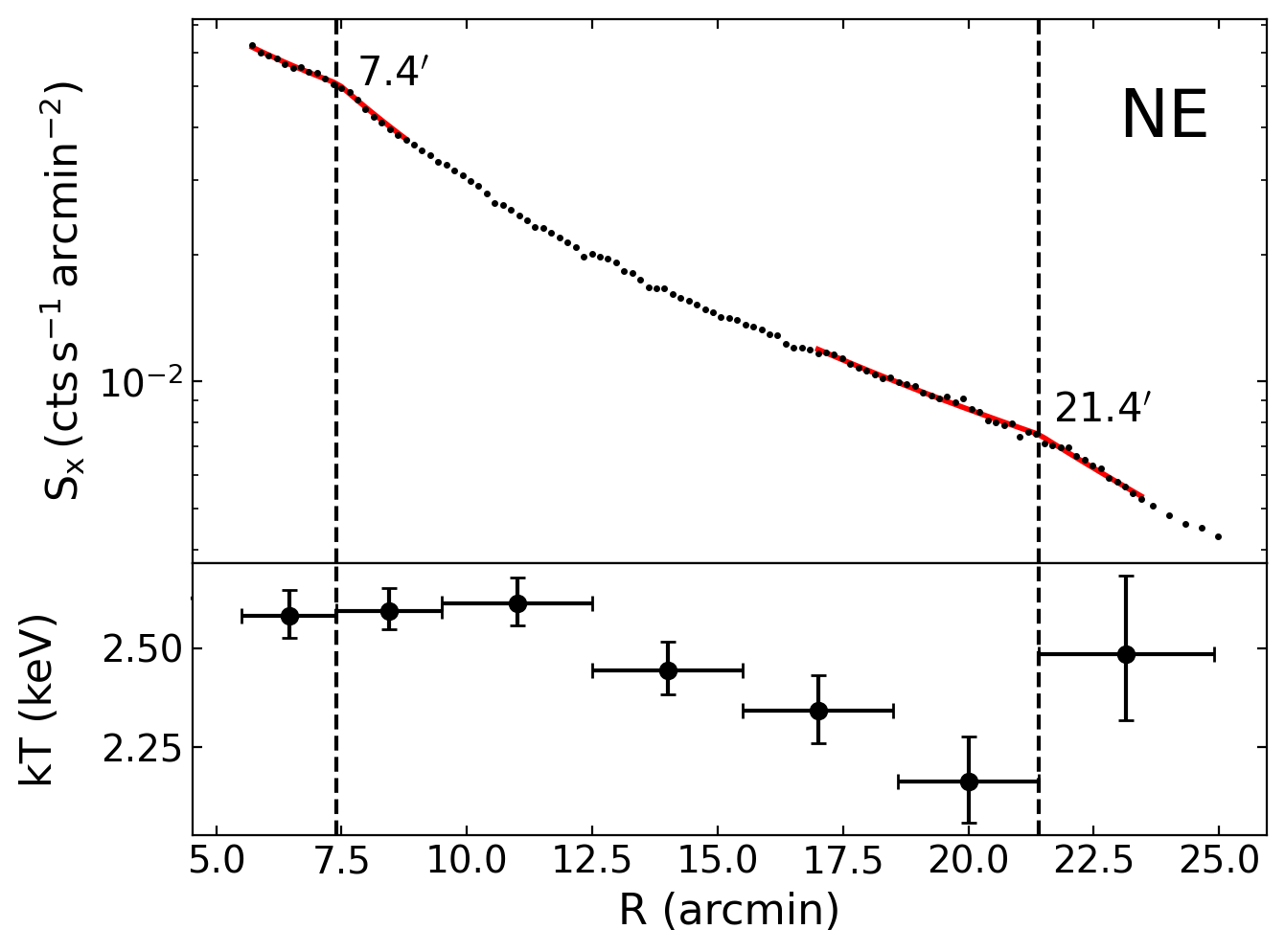}
\includegraphics[width=0.49\linewidth]{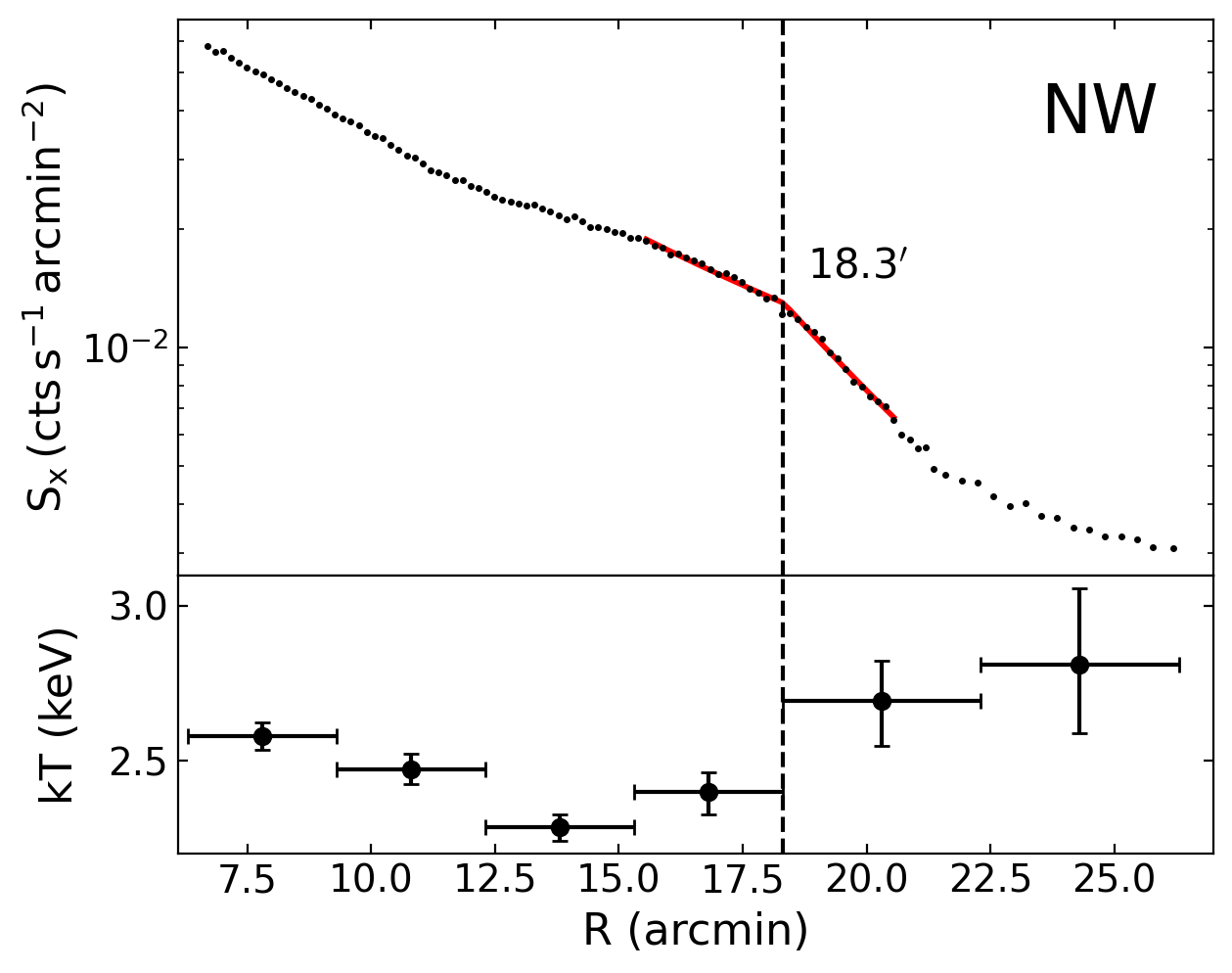}
\includegraphics[width=0.49\linewidth]{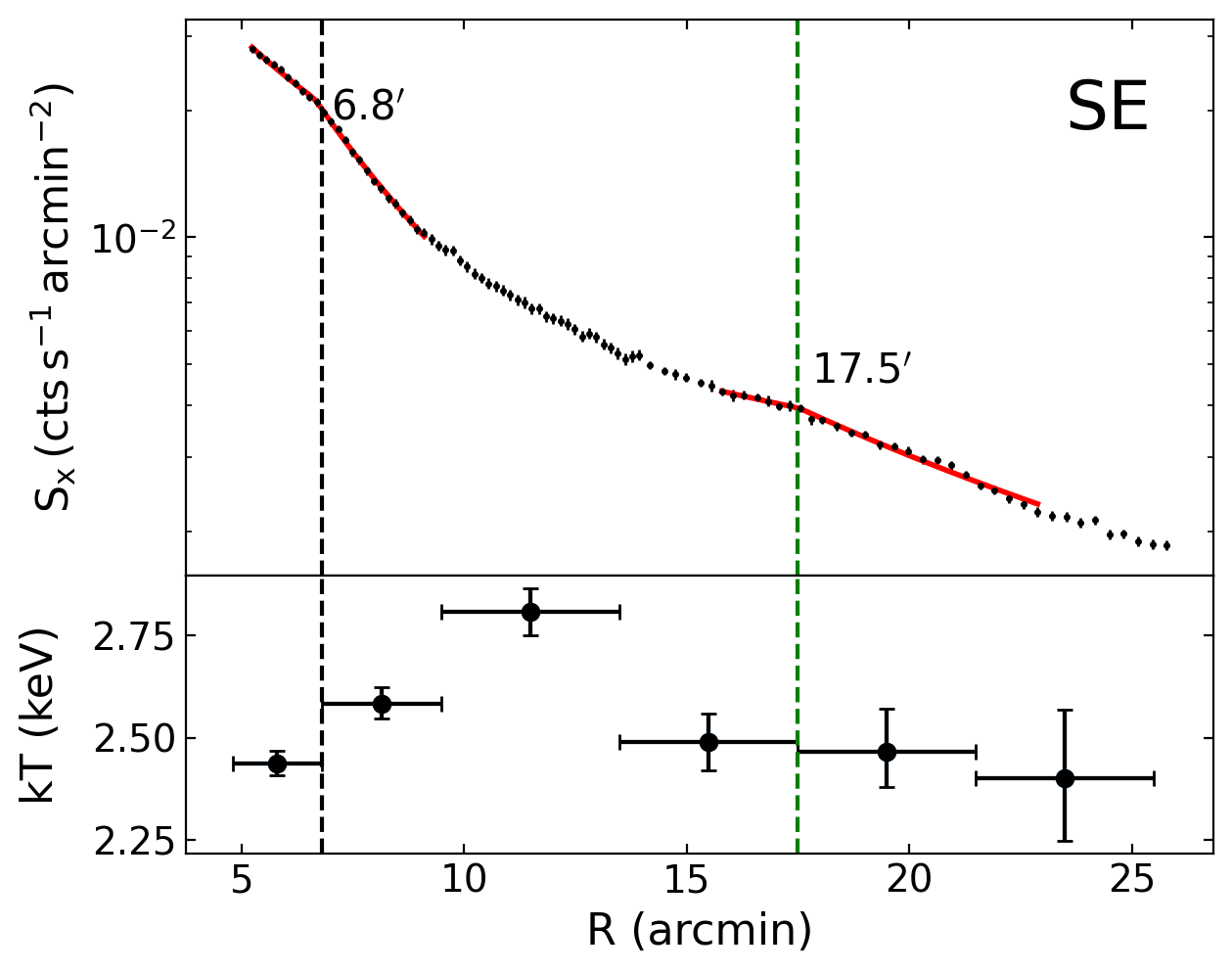}
\includegraphics[width=0.49\linewidth]{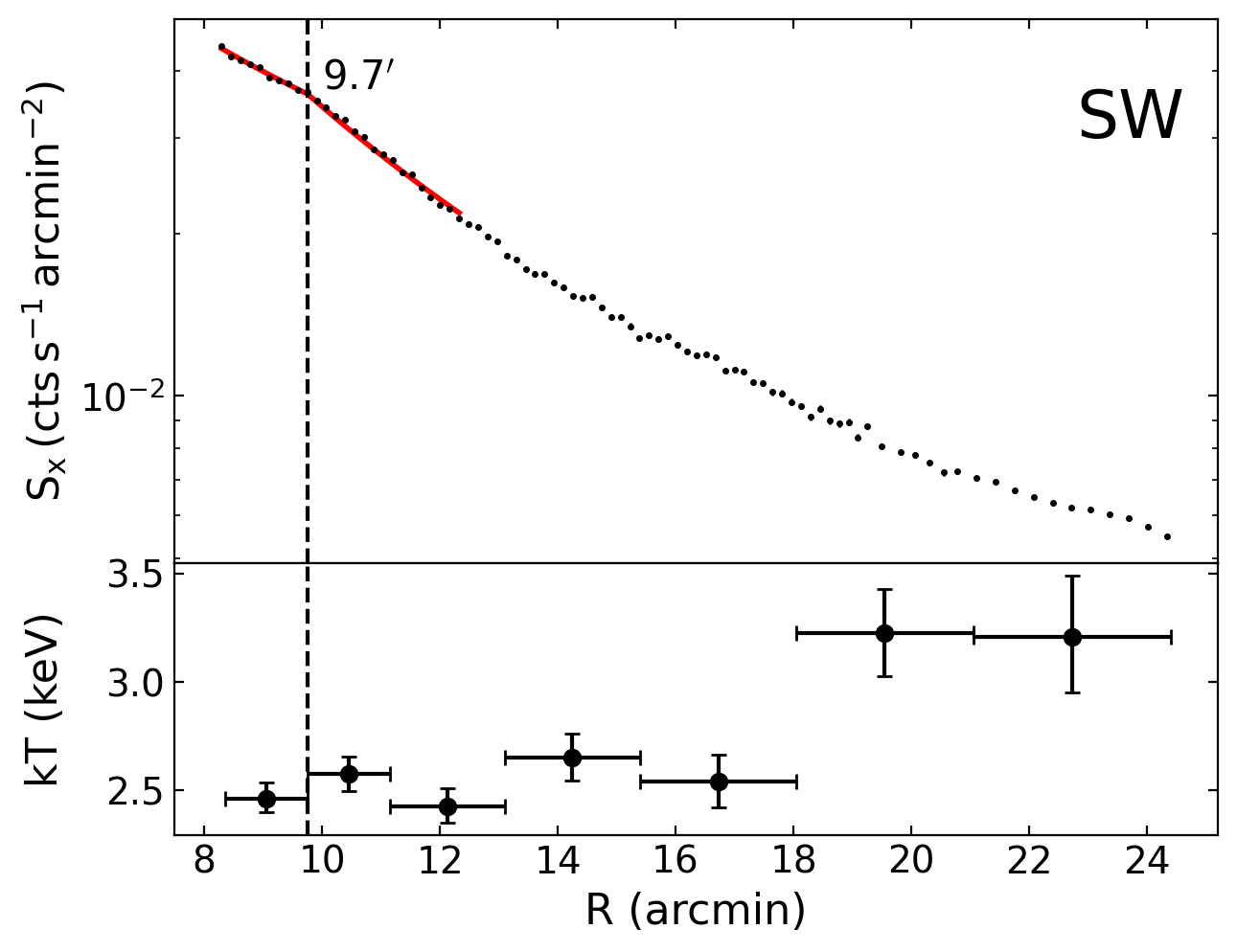}

\caption{Radial profiles of surface brightness and temperature in the four sectorial regions marked NE, NW, SE, and SW. The surface brightness was extracted from Fig.~\ref{fig:full}. The red curves indicate the best-fit broken power-law model. The temperature increases and decreases at density discontinuities are marked by vertical black and green lines, respectively.
}
\label{fig:sector_radial}
\end{figure*}

In the residual map (see Fig.~\ref{fig:res}), we defined four sector regions of interest (NW, SE, NE, and SW). 
Using Fig.~\ref{fig:full}, we extracted the radial profiles of the surface brightness and temperature in each region (see Fig.~\ref{fig:sector_radial}). The surface brightness profiles near the cold fronts were fit with a broken power law. 
The best‐fit parameters of the broken power‐law model are listed in Table~\ref{tab:bpl}.

In the northeast, two surface brightness drops are detected at $R = 7.44 \pm 0.04^{\prime}$ and $R = 21.40 \pm 0.14^{\prime}$. At $R = 21.40 \pm 0.14^{\prime}$, a modest increase in the temperature is found from $2.16^{+0.11}_{-0.10}~\mathrm{keV}$ to $2.50^{+0.19}_{-0.16}~\mathrm{keV}$ at the $1.7\sigma$ level (the uncertainties are quoted at the $1\sigma$ level). 
While the temperature change alone is not highly significant, the corresponding surface brightness discontinuity and the location in the outskirts of M87 (where the S/N is lower) suggest that this feature likely marks the outer boundary of the northeastern spiral structure.

In the northwest, a drop in the surface brightness is found at $R = 18.33 \pm 0.08^{\prime}$, accompanied by a rise in temperature from $2.39^{+0.06}_{-0.07}$ to $2.70^{+0.12}_{-0.14}~\mathrm{keV}$, corresponding to a significance of approximately $2.2\sigma$.
Despite the moderate significance of the temperature jump, the surface brightness edge, coupled with the low S/N in the outskirts, supports the classification of this feature as a cold front. This agrees with previous findings.

In the southeast, a cold front is identified at $R = 6.76 \pm 0.03^{\prime}$, indicated by a drop in the surface brightness and a temperature increase from $2.43^{+0.03}_{-0.02}$ to $2.58^{+0.04}_{-0.03}~\mathrm{keV}$ at the $3.3\sigma$ level. At $R = 17.52 \pm 0.22^{\prime}$, the surface brightness shows a mild discontinuity, transitioning from a deficit to an excess in the residual map (see Fig.~\ref{fig:res}). This feature may correspond to the inner boundary of the southeastern portion of the spiral structure.

A region of excess surface brightness is detected in the southwest at $R = 9.75 \pm 0.07^{\prime}$, which might be associated with an arc-shaped feature from past AGN activity \citep{forman2003reflections}. The surface brightness discontinuity is described by a broken power-law model, and a corresponding radial temperature increase further supports the possible presence of a weak cold front.

We compared the radial temperature profile derived from \textit{EP-FXT} observations with the results reported by \citet{gatuzz2022measuring} based on \textit{XMM-Newton} data. The two profiles show a broadly consistent trend, with comparable shapes and temperatures ranging from approximately 2 to 3 keV.

\begin{table*}[t]
\centering
\footnotesize
\caption{
Parameters of the best-fit broken power-law model.}
\label{tab:bpl}
\tabcolsep=10pt
\begin{tabular*}{0.8\textwidth}{ccccc}
\hline\hline
region & $R_{\rm break}$ (arcmin)
 & $I_{0}$ ( $10^{-2}$ cts\,s$^{-1}$\,arcmin$^{-2}$)
 & $\alpha_{1}$
 & $\alpha_{2}$ \\
\midrule
NW & $18.33 \pm 0.08$ & $1.29 \pm 0.02$ & $2.27 \pm 0.08$ & $5.85 \pm 0.23$\\
\hline
SE & $6.76 \pm 0.03$ & $2.13 \pm 0.02$ & $1.18 \pm 0.03$ & $2.39 \pm 0.03$ \\
 & $17.52 \pm 0.22$ & $0.39 \pm 0.01$& $0.88 \pm 0.17$ & $1.96 \pm 0.05$ \\
\hline
SW & $9.75 \pm 0.07$ & $3.62 \pm 0.05$ & $1.19 \pm 0.06$ & $2.16 \pm 0.01$ \\
\hline
NE & $7.44 \pm 0.04$ & $5.07 \pm 0.05$ & $0.76 \pm 0.03$ & $1.80 \pm 0.01$ \\
 &$21.40 \pm 0.14$& $0.75 \pm 0.01$& $2.01 \pm 0.04$& $3.70 \pm 0.14$ \\
\hline
\end{tabular*}
\tablefoot{
Best-fit parameters of the broken power-law model to the surface brightness profile, with uncertainties quoted at the 1$\sigma$ confidence level. The model includes the break radius ($R_{\rm break}$), which marks the location of the surface brightness discontinuity; the inner and outer power-law indices ($\alpha_{1}$ and $\alpha_{2}$), which describe the slope of the profile on either side of the break; and the normalization constant ($I_{0}$), which represents the surface brightness at the break radius.
}
\end{table*}

\section{Discussion}\label{sec:5}

The imaging and imaging-spectral analyses revealed detailed structures at the center of the Virgo cluster.
In addition to the well-known X-ray arms, cavities, and cold fronts, a large-scale sloshing spiral pattern is identified in the GGM and residual images, where it connects the two previously recognized cold fronts.
The low particle background and wide FoV allow the \textit{EP-FXT} observations to provide a continuous view of the spiral structure in the outskirts of M87. This is consistent with a similar pattern seen in the mosaic image from \textit{XMM-Newton} \citep{simionescu2010metal}.
Building on these observations, we employed \textit{EP-FXT} data to construct two-dimensional thermodynamic maps of the outskirts of M87 to provide further insight into the large-scale physical properties of the spiral structure.
In this section, we discuss the properties of the spiral structure and compare the results with those obtained by \textit{XMM-Newton}.

\subsection{Comparison with XMM-Newton}

\begin{figure*}[tbh]
\centering
\includegraphics[width=0.49\linewidth]{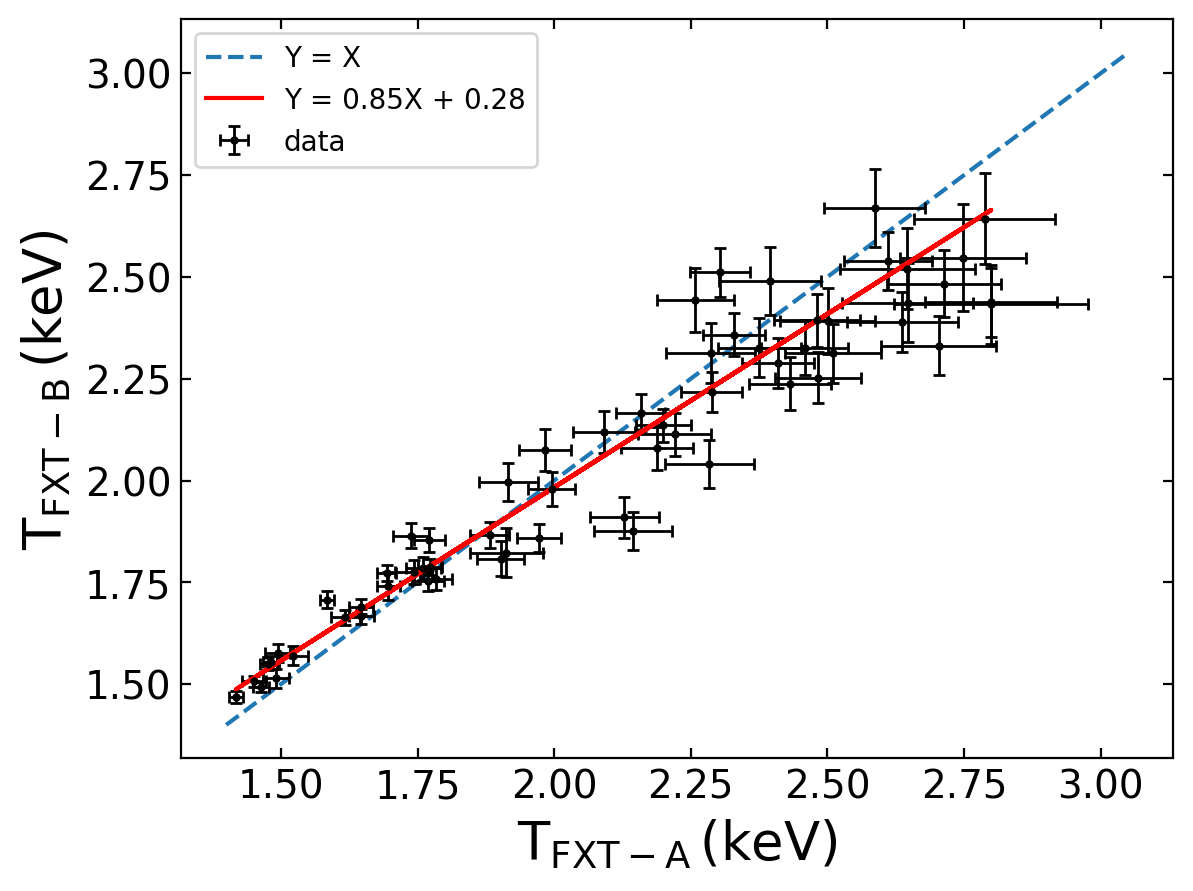}
\includegraphics[width=0.49\linewidth]{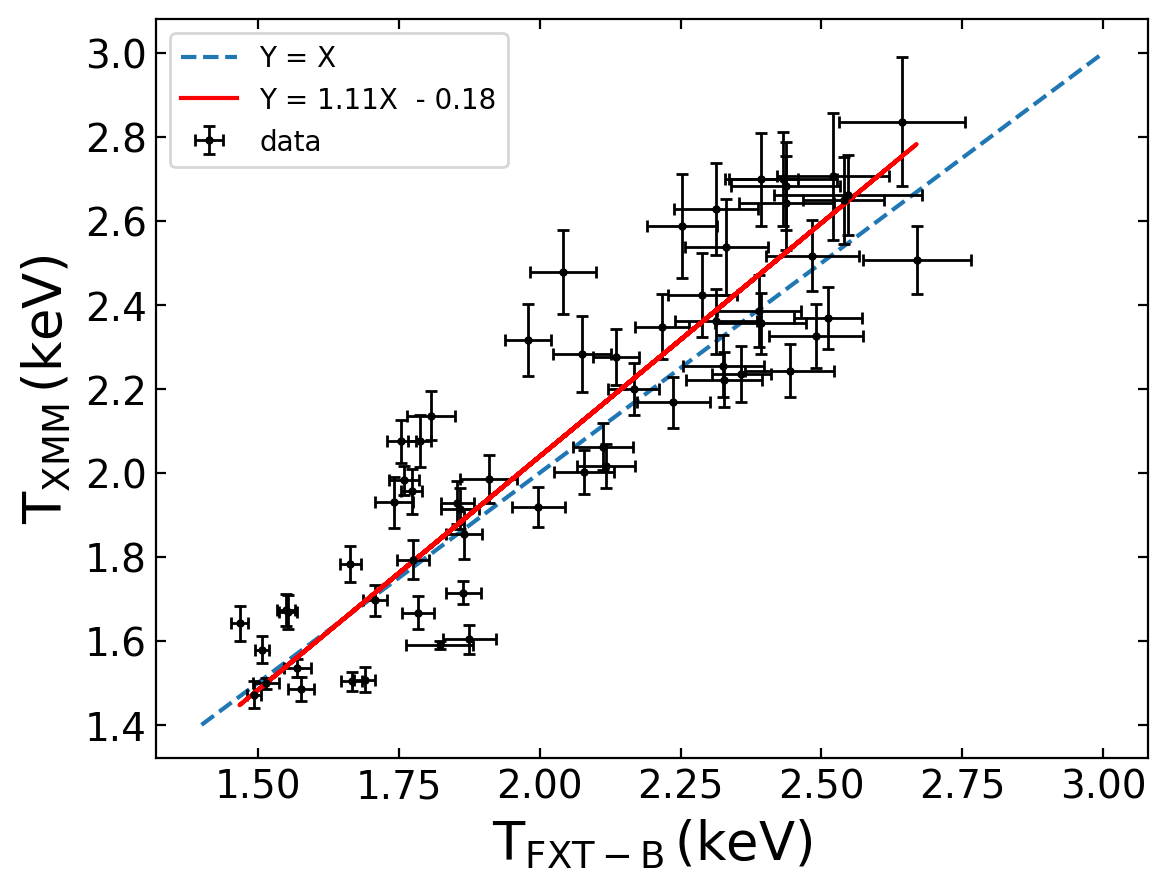}
\caption{Comparison of the temperature in each pixel between \textit{EP-FXT} and \textit{XMM-Newton}. The dashed blue line represents the 1:1 ratio. The red line represents the best-fit linear regression, shown in the legend.}
\label{fig:cmp}
\end{figure*}

Using the same \texttt{contour-binning} regions as defined for the \textit{EP-FXT} data, we performed spectral fitting on the archival \textit{XMM-Newton} observation (ObsID 0803670601) over the 0.5–7.0 keV energy band, adopting the same model described in Sect.~\ref{sec:4}.
The temperatures for each region were compared with those obtained from \textit{EP-FXT}, as shown in Fig.~\ref{fig:cmp} and Table~\ref{tab:cmp}.
In each panel, the blue dashed line denotes the one-to-one correspondence, while the red solid line represents the best-fit linear regression.

As shown in Fig.~\ref{fig:cmp}, the temperatures measured by FXT-A and FXT-B are consistent within the uncertainties, although the temperature from FXT-A appears to be slightly higher than that from FXT-B.
This discrepancy may be attributed to the greater effective area of FXT-B at lower energies.
Overall, the temperatures obtained from \textit{EP-FXT} generally agree with those derived from \textit{XMM-Newton}, despite minor discrepancies.
These differences may result from the multitemperature components within the ICM and from calibration differences of the two telescopes, which together can lead to systematic discrepancies in the temperature measurements \citep{migkas2024srg,liu2023x,schellenberger2015xmm}.
Overall, \textit{EP-FXT} delivers results that are consistent with and comparable to those of \textit{XMM-Newton}.

\begin{table*}[t]
\centering
\footnotesize
\caption{Comparison of the temperatures measured by \textit{EP-FXT} and \textit{XMM-Newton}.}
\label{tab:obs}
\tabcolsep 12pt 
\begin{tabular*}{0.7\textwidth}{cccc}
\hline\hline
   Instruments X-Y & Energy band (keV) & a & b\\
  
    \hline
     FXTA-FXTB & 0.5-7.0 & $0.85\pm0.03$ &  $0.28\pm0.04$  \\
    FXTA-XMM & 0.5-7.0 & $1.11\pm0.07$ & $-0.18\pm0.13$  \\
\hline
\label{tab:cmp}
\end{tabular*}
\tablefoot{Comparison of temperature fits for different instruments X and Y. Here, a and b are the linear fitting parameters.}
\end{table*}

\subsection{Giant sloshing spiral}

Based on the above analysis, we further confirmed the previously identified spiral structure using the GGM and residual images from \textit{EP-FXT}, which provide a comprehensive view of the feature.
Subsequently, we performed an imaging-spectrum analysis to construct two-dimensional maps of the temperature, metallicity, and entropy in the outskirts of M87, providing a spatially resolved characterization of the thermodynamic properties of the spiral structure.
The results show that throughout the spiral, the metallicity is enhanced, while temperature and entropy are reduced.
In addition, the spiral structure exhibits significant surface brightness enhancements. Cooler gas is preferentially concentrated in the brighter regions, where it forms an alternating pattern of bright cold areas and dimmer hotter regions.
Numerical simulations by \citet{roediger2011gas} predicted that these structures should exhibit low temperatures, high densities, and an elevated metallicity.
Our observations are consistent with these predictions and support the scenario in which the large-scale spiral structure is produced by gas sloshing that was triggered by a minor merger event \citep{ascasibar2006origin}.

Figure~\ref{fig:res_img} shows the large spiral structure in Virgo as detected by \textit{EP-FXT}. It consists of an inner ring-like structure and outer spiral arms.
Northeast of the spiral structure (as indicated by the white arrow), the X-ray surface brightness is lower than in the surrounding regions and forms a concave feature. Two additional X-ray concave regions are also present in the west-southwest. 
This structure is similar to the Kelvin-Helmholtz rolls presented in the simulations and the smeared-out spiral structures observed in A496 \citep{walker2017there,ghizzardi2014metal}. This similarity may be attributed to the influence of the Kelvin-Helmholtz instability, which generates turbulence and disturbances near the cold front that might affect the formation of the spiral structure.

Overall, \textit{EP-FXT} provides a clear and comprehensive view of the large-scale spiral pattern in M87. The spiral feature appears to rotate counterclockwise and extends from the cold front in the southeast to the cold front in the northwest. It further spirals toward the cold gas clump in the southern outskirts. This spiral feature is associated with a low temperature, low entropy, and high metallicity, consistent with expectations for gas-sloshing phenomena. The prominent and well-defined morphology revealed by \textit{EP-FXT} offers valuable insights into the formation and evolution of spiral structures in galaxy clusters.

\section{Conclusion}\label{sec:6}

We provided a detailed analysis of the images and spectra of the Virgo cluster with \textit{EP-FXT}.
The images show a prominent and complete large-scale spiral structure.
Furthermore, we produced two-dimensional maps of the temperature, entropy, and metallicity that clearly illustrated the characteristics of the M87 spiral structure. The main results of this work are summarized below.

\begin{enumerate} 

\item In the GGM image, multiple structures are identified. Two prominent X-ray arms are clearly visible: The outer side of the southwest arm bends toward the southeast, and the east arm bends toward the northwest. A shock from the northwest appears to be connected to the east arm. Furthermore, cold fronts in the southeast and northwest are also readily discernible. These structures are broadly consistent with previous findings.

\item Based on the residual map (see Fig.~\ref{fig:res}), a ring-like structure with an enhanced surface brightness is found in the outer X-ray arms. This ring-like structure is connected to the northwestern cold front and rotates counterclockwise. It forms a large, outward-extending spiral structure. This structure has a low temperature, low entropy, and high density. These characteristics are consistent with gas sloshing induced by the merger of a gasless subcluster.

\item The temperatures measured with \textit{EP-FXT} are consistent with those obtained with \textit{XMM-Newton} within the errors.

\begin{figure}
\centering
\includegraphics[width=\linewidth]{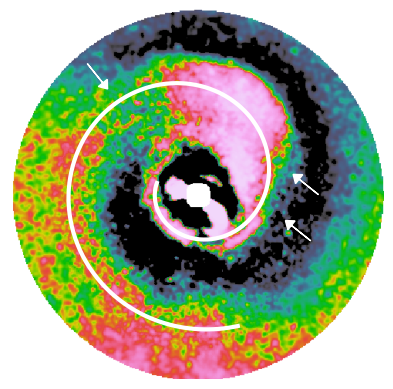}
\caption{\textit{EP-FXT} surface brightness residual image in the 0.3–2.5 keV
band showing a spiral structure. The white arrowheads indicate locations where
this structure is truncated, possibly because of the Kelvin-Helmholtz instability. The central ring
structure of M87 and the outer spiral structure appear to be part of a single
larger structure, as delineated by the white spiral line.}
\label{fig:res_img}
\end{figure}
\end{enumerate}

\begin{acknowledgements}
\textit{EP} is a space mission supported by
Strategic Priority Program on Space Science of Chinese Academy
of Sciences, in collaboration with ESA, MPE and CNES (Grant
No.XDA15310303, No.XDA15310103, No.XDA15052100).
This work is supported by the International Partnership Program of Chinese Academy of Sciences，Grant No. 013GJHZ2024015FN.
CJ acknowledges the National Natural Science Foundation of China through grant 12473016, and the support by the Strategic Priority Research Program of the Chinese Academy of Sciences (Grant No. XDB0550200).
A.L. acknowledges the supports from the National Natural Science Foundation of China (Grant No. 12588202). A.L. is supported by the China Manned Space Program with grant no. CMS-CSST-2025-A04.
\end{acknowledgements}

  \bibliographystyle{aa} 
  \bibliography{main}

\end{document}